\documentclass{llncs}

\usepackage[utf8]{inputenc}

\usepackage{graphics} 
\usepackage{epsfig} 
\usepackage{mathptmx} 
\usepackage{times} 
\usepackage{amsmath} 
\usepackage{amssymb}  
\usepackage{multirow}
\usepackage{lineno}
\usepackage{tikzpagenodes}
\usepackage{multirow}
\usepackage{bigdelim}
\usepackage{subfigure}
\usepackage{float}

\usepackage[dvips]{rotating} 
\usepackage{array}
\usepackage{multirow}
\usepackage{subfigure}
\usepackage{rotating}

\usepackage{textcomp}
\usepackage{epsfig}
\usepackage{changes}

\title{Frozen-to-Paraffin: Categorization of Histological Frozen Sections by the Aid of Paraffin Sections and Generative Adversarial Networks}

\author{****\\****}
\author{
Michael Gadermayr$^1$ \and Maximilian Tschuchnig$^1$ \and Lea Maria Stangassinger$^2$ \and \\ Christina Kreutzer$^3$ \and Sebastien Couillard-Despres$^3$ \and \\ Gertie Janneke Oostingh$^2$ \and Anton Hittmair$^4$}

\institute{****\\****}
\institute{
$^1$ Salzburg University of Applied Sciences, Department of Information Technology and Systems Management\\
$^2$ Salzburg University of Applied Sciences, Department of Biomedical Sciences\\
$^3$ Spinal Cord Injury and Tissue Regeneration Center Salzburg, Research Institute of Experimental Neuroregeneration\\
$^4$ Kardinal Schwarzenberg Klinikum, Department of Pathology and Microbiology
}

\begin{document}

\maketitle

\begin{abstract}
	In contrast to paraffin sections, frozen sections can be quickly generated during surgical interventions.
	This procedure allows surgeons to wait for histological findings during the intervention to base intra-operative decisions on the outcome of the histology.
	However, compared to paraffin sections, the quality of frozen sections is typically lower, leading to a higher ratio of miss-classification.
	In this work, we investigated the effect of the section type on automated decision support approaches for classification of thyroid cancer.
	This was enabled by a data set consisting of pairs of sections for individual patients.
	Moreover, we investigated, whether a frozen-to-paraffin translation could help to optimize classification scores. Finally, we propose a specific data augmentation strategy to deal with a small amount of training data and to increase classification accuracy even further.
\end{abstract}

\keywords{Histology \and Frozen sections \and Generative adversarial networks \and Thyroid cancer \and Data augmentation \and Whole slide image classification}

\section{Motivation}\label{sec:introduction}
Whole slide imaging is capable of effectively digitizing specimen slides, showing both the microscopic detail and the larger context, without any significant manual effort.
Visual pathological examination of slides, independent whether they are digitized or not, is time-consuming and error prone, due to the large amount of information available in the highly-resolved image data. 
These facts provide the incentive for the development of automated tools to support pathologists during clinical routine. Recently, automated segmentation~\cite{myHalicek19a,Gadermayr19b}, normalization~\cite{myBentaieb18a} and classification approaches~\cite{myHou16a} have been proposed, mainly based on state-of-the-art deep learning approaches.

\begin{figure}
    \centering
    \includegraphics[width=\linewidth]{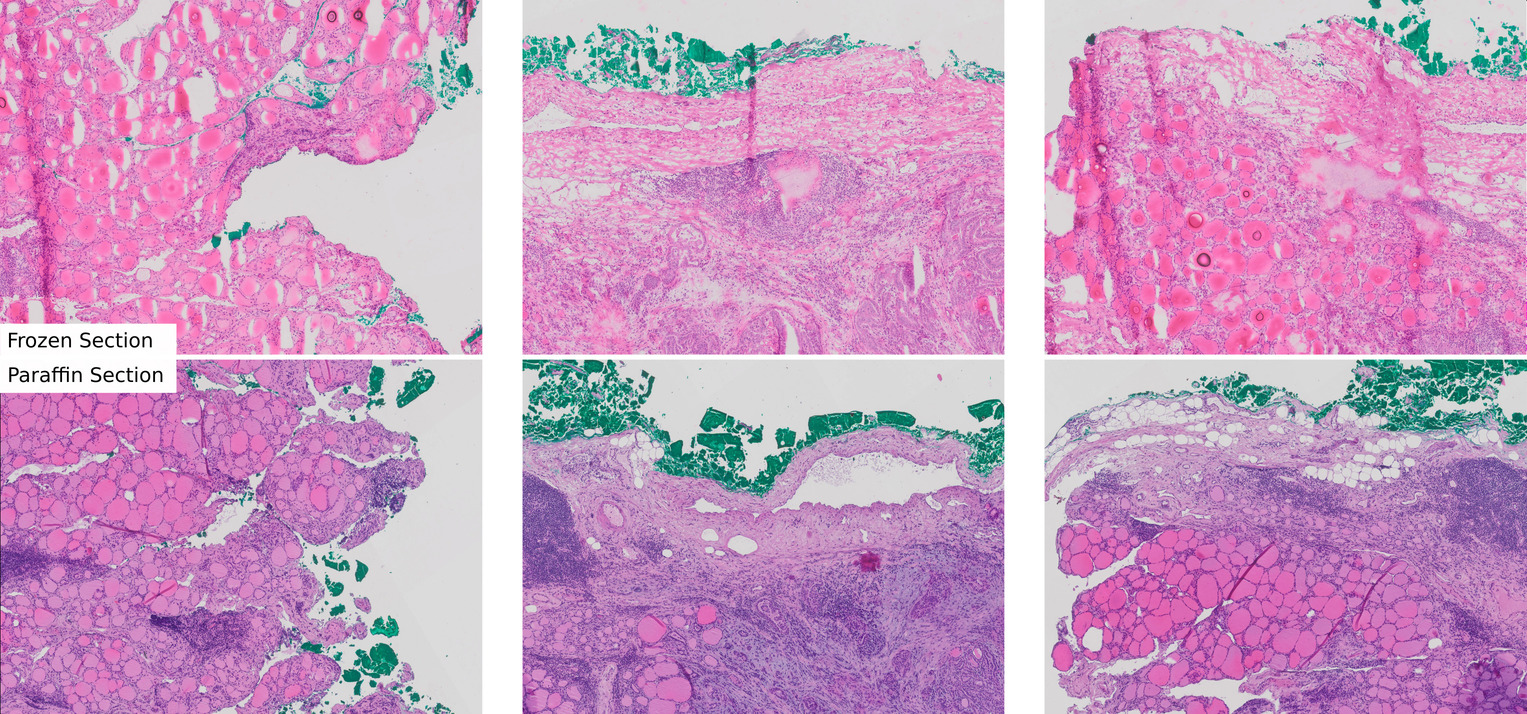}
    \caption{Semi-corresponding tissue showing frozen sections (top row) and paraffin embedded sections (bottom row). The two preparations were obtained from neighboring tissue (and thereby show similarities).}
    \label{fig:example_corresponding}
\end{figure}

Tissue embedding is the first step in the histological pipeline for tissue preparation.
In the case of paraffin sections, the tissues are embedded in a solid medium both to support the tissue and to enable the cutting of thin tissue slices.
Paraffin embedding is probably the most commonly used technique (Fig.~\ref{fig:example_corresponding}, bottom row) that is compatible with a large variety of staining methods and allows thin sectioning (down to few micrometers) with a high visual quality.
So-called frozen sections (Fig.~\ref{fig:example_corresponding}, top row), are typically generated during interventions (e.g. cancer resections) in order to achieve information on malignancy as quick as possible, because the preparation time for paraffin sections is typically too long to be used for that purpose.
Frozen sections therefore allow surgeons to wait during interventions for the histological examination in order to base further procedures on the outcome.
However, compared to paraffin sections, the quality of frozen sections is typically lower, leading to a higher ratio of miss-classification when performed by clinical experts~\cite{myNajah19a,myOsamura08a,myHuber07a}.
The cellular structure can be seen more clearly in paraffin sections, as these are fixed before embedding in paraffin. This is not the case in frozen sections with the result of partly indiscernible or damaged tissue features~\cite{myLeteurtre01a,myUdelsman01a}.

Generative adversarial networks, recently proved to facilitate a translation between medical imaging domains, such as CT to MRI~\cite{myWolterink17a} or translations between different stainings~\cite{Gadermayr19b}, based on unpaired image data only. A limitation in this application scenario is often the fact that so-called one-to-many mappings~\cite{myAlmahairi18a} complicate training of cyclic architectures~\cite{myZhu17a,myYi17a}. The challenge in many application scenarios is, that a characteristic in one domain often can be mapped to more than one pendent in the other modality. This can lead to conflicts when using pixel-based losses~\cite{myZhu17a,myYi17a}. Even more recently, methods-of-resolution were proposed, tackling also highly challenging translation settings by adding an additional latent variable~\cite{myLiu17b,myAlmahairi18a,myHuang18a} or by replacing the pixel-based cycle-consistency loss with a feature-based loss~\cite{myPark20a}.

\subsection*{Contribution} 
The contribution of this work is manifold:
Based on a generic fully automated pipeline, we first investigated the effect of the section type on automated decision support approaches for the classification of thyroid cancer. A unique data set showing similar tissue from both modalities, allowed for an unbiased evaluation.
Secondly, we investigated, whether a frozen-to-paraffin conversion can help to increase classification scores of frozen sections. To improve the performance in the case of small data even further, a specific data augmentation strategy is proposed.

\begin{figure}[tb]
	\includegraphics[width=\linewidth]{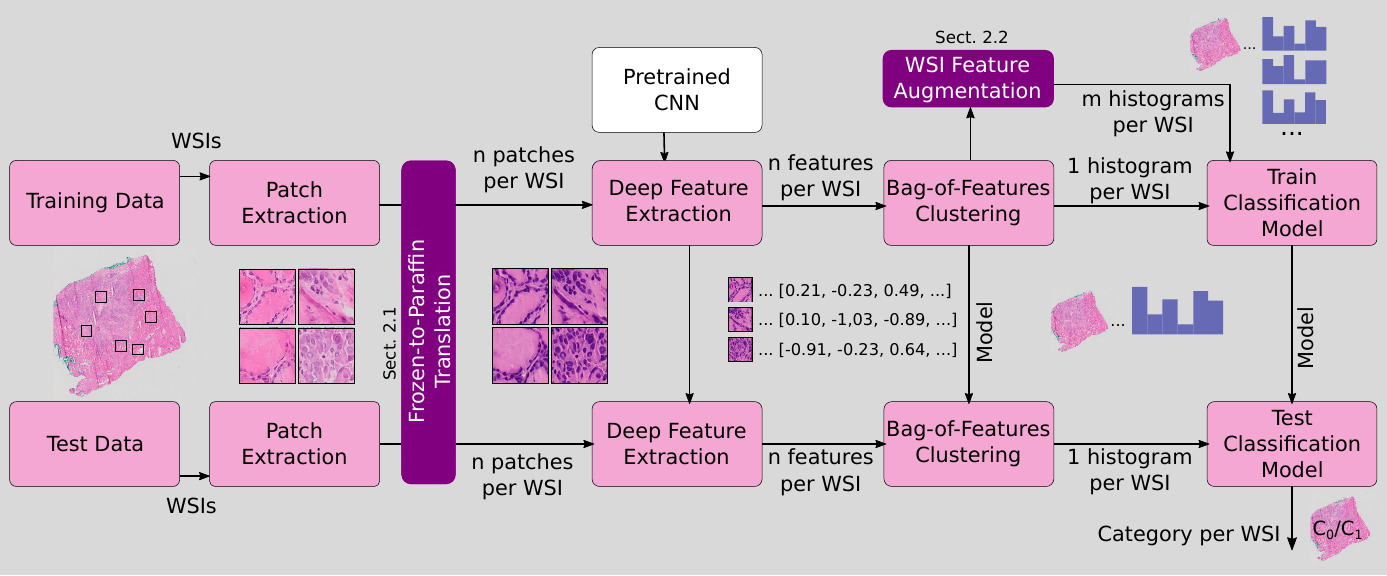}
	\caption{Outline of the proposed basic pipeline with the proposed additional stages (Frozen-to-Paraffin Translation \& WSI Feature Augmentation) for improving accuracy of frozen sections. Input of the pipeline is a set of whole slide images for training and one for testing. Output is a classification label for each test image.}
	\label{fig:outline}
\end{figure}

\section{Methods}

To investigate the impact of the section type and section-translations on the classification accuracy of automated approaches, we rely on a generic straight-forward patch-based whole slide image classification architecture (Fig.~\ref{fig:outline}), similar to~\cite{myHou16a,myDimitriou19a}. As a holistic classification of whole slide images is not effective~\cite{myHou16a}, in a first step patches are randomly extracted in image areas showing mainly tissue.
These patches are fed into a pre-trained convolutional neural network (CNN) for means of feature extraction. The $n$ feature vectors per whole slide image, obtained from training data, are then clustered (k-means) and aggregated. This aggregation of the clustered patches leads to a single histogram per whole slide image (Bag-of-Features). With these histograms, together with the whole slide image labels, a classification model (SVM) is finally trained. For testing purpose, both, the bag-of-features approach and the final classification model is optimized with training data, whereas the pre-trained CNN is not further adapted. 
This generic architecture is preferred to a more specific feature learning-based approach, due to the small number of available images. Moreover, the training efficiency allows to investigate a large number of settings combined with a large number of random splits (overall, we trained the pipeline 768 times) to facilitate general statements.

We propose two adjustments: firstly, we propose to apply frozen-to-paraffin translation to optimize the quality of frozen sections, by means of a state-of-the-art image-translation approach (Sect.~\ref{sec:i2i}). Secondly, we propose to augment the available training samples (per whole slide image), by a very efficient sampling strategy (Sect.~\ref{sec:augment}). This adjustment allows the simulation of a larger amount of available training data (even though a larger data set would be advantageous) to make more general statements on the effect of image-translation.


\subsection{Frozen-to-Paraffin Translation} \label{sec:i2i}
For frozen-to-paraffin translation, a generative adversarial network based on contrastive learning~\cite{myPark20a} is employed, which solves the problem of one-to-many mappings of cyclic-architectures~\cite{myZhu17a} by means of a feature based loss. 

We denote the domain of frozen sections as $F$ ($f \in F$) and the domain of paraffin sections as $P$ ($p \in P$). We trained the generator $G$ consisting of an encoder $G_{enc}$ and a decoder $G_{dec}$ to perform the mapping $\hat{p} = G_{dec}(G_{enc}(f))$ were $\hat{p}$ represents the virtual paraffin pendant of the original frozen section $f$.
We optimized a loss criterion consisting of a weighted sum of a GAN loss $\mathcal{L}_{GAN}$, a patch similarity loss $\mathcal{L}_{PatchNCE}(G,H,F)$ forcing corresponding patches to share content and an additional regularization term $\mathcal{L}_{PatchNCE}(G,H,P)$.
In summary, the loss can be formulated as follows
\begin{equation}
L = \mathcal{L}_{GAN}(G,D,F,P) + \lambda_F \mathcal{L}_{PatchNCE}(G,H,F) + \lambda_G \mathcal{L}_{PatchNCE}(G,H,P) \; ,
\end{equation}
with $D$ being the discriminator, $H$ being a two-layer perceptron and $\lambda_F$ and $\lambda_G$ being scalar weights. The patch similarity loss was computed on features level, i.e. it was computed after the encoder part of the generator ($G_{enc}$) was applied to the image.
We utilized a ResNet with nine blocks as the generator, patchGAN~\cite{myIsola16a} as the discriminator, a least squares GAN loss, a batch size of one and Adam as an optimizer with an initial learning rate of 0.002. The model was trained  for  40 epochs. $\lambda_F$ and $\lambda_G$ were both set to 1.0.
For details, we refer to~\cite{myPark20a}. Finally, we used the Pytorch reference implementation with the default 'cut' setting.

\subsection{WSI Feature Augmentation} \label{sec:augment}

Due to the rather small amount of available whole slide images, we incorporated feature-augmentation to obtain several features for one whole slide image.
In the patch extraction stage, for each original image, 512 patches with a size of $256 \times 256$ pixels were randomly (uniformly) selected from regions showing (at least 75 \%) tissue. In the bag-of-features stage, each patch was assigned to a cluster center. The final feature vector was represented by the distribution of patches with respect to these cluster centers. In the conventional setting, this generates exactly one feature vector for each whole slide image.

To increase the number of feature vectors for training, we reduced the number of sampled patches to a certain ratio ($r_{patch}$) of the 512 originally extracted patches and repeated this process eight times. Thereby, we achieved a clearly increased number of features ($8 \times$) without high computational or manual effort. Obviously, the features created within one whole slide images showed a higher correlation than features between images. However, the motivation for this procedure was provided by the patchy distribution of relevant disease markers. The generation of several random samples can help to properly fill the feature space before training a classification model. In the experimental evaluation, two different sampling strategies (Aug1: $r_{patch} = 75 \%$, Aug2: $r_{patch} = 50 \%$) were evaluated.

\subsection{Data Set} \label{sec:dataset}
In this work, we aimed at distinguishing different nodular lesions of the thyroid, focusing especially on indeterminate follicular lesions and papillary carcinoma. This differentiation is crucial, due to different treatment options, in particular with respect to the extent of surgical resection of the thyroid gland.
The data set consisted of 80 whole slide images overall, i.e. 40 slides were available for each section type. All images were acquired during clinical routine at the 
Kardinal Schwarzenberg 
Hospital. They were labeled by an expert pathologist with over 20 years experience. A total of 42 (21 per modality) slides were labeled as papillary carcinoma while 38 (19 per modality) were labeled as follicular carcinoma.
During the generation of data, the focus was placed on extracting samples showing similar tissue (for examples, see Fig.~\ref{fig:example_corresponding}). Perfect matching was not possible , since the tissue blocks needed to be separated and individually processed for each modality. As a result, two consecutive slides showing the perfect immediately neighboring tissue areas~\cite{Gupta18a} cannot be obtained for this procedure. Nevertheless, the slides showed similar underlying content, which is an important criterion when comparing classification accuracy for both modalities.
However, due to the imperfect alignment of the data, paired image translation approaches were not applicable.
For frozen sections, fresh tissue was frozen at $-15^\circ$ Celsius, slides were cut (thickness $5 \mu m$) and stained immediately with hematoxylin and eosin.
For paraffin sections, tissue was fixed in 4\% phosphate-buffered formalin for 24 hours. Subsequently formalin fixed paraffin embedded tissue was cut (thickness $2 \mu m$) and stained with hematoxylin and eosin.
Images were digitized with an Olympus VS120-LD100 slide loader system. Overviews at a 2x magnification were generated to manually define scan areas, focus points were automatically defined and adapted if needed. Scans were performed with a 20x objective. Image files were stored in the Oympus vsi format based on lossless compression.

\subsection{Experimental Details} \label{sec:setup}
The available data was randomly separated into training (80 \%) and test data (20 \%). The whole pipeline, including the separation was repeated 32 times, in order to achieve representative scores. Due to the almost balanced setting (see Sect.~\ref{sec:dataset}), the classification accuracies (mean and standard deviation) were finally reported.

\paragraph{Patch Extraction}
Patches were randomly extracted from the whole slide image, based on uniform sampling. For each patch, we checked that at least 75 \% of the area was covered with tissue in order to exclude empty areas. We extracted a total of 512 patches with a size of $256 \times 256$ pixel per whole slide image.

\paragraph{Deep Feature Extraction}
Due to the limited number of whole slide images, we did not train the feature extraction stage~\cite{myHou16a}, but used a pre-trained network instead. Specifically, we use a ResNet18 pre-trained on the image-net challenge data, due to the high performance in previous work on similar data~\cite{myDimitriou19a}.
ResNet18 is particularly appropriate due to the rather low dimensional output of the convolutional layers (512 neurons). We directly forward the convolutional layers' output to the next stage.

\begin{figure}[tb]
	
	\includegraphics[width=\linewidth]{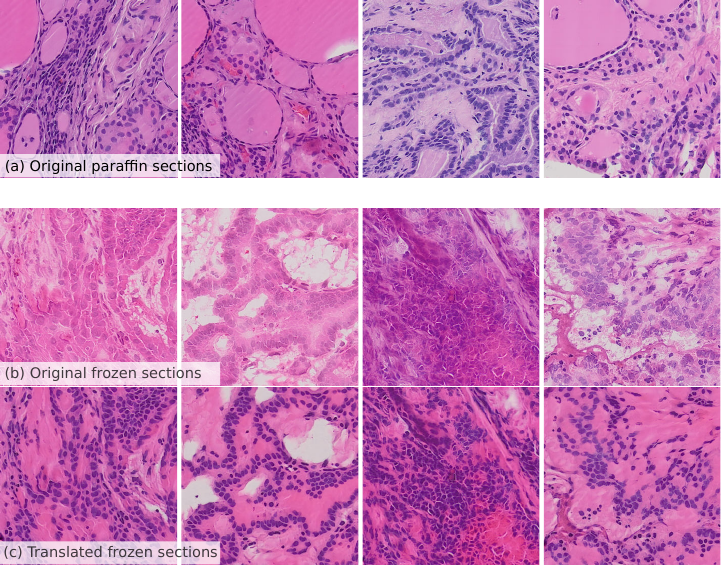}
	%
	\caption{Example patches with a size of $700 \times 700$ pixels showing (a) paraffin sections, (b) frozen sections and (c) the corresponding 'frozen-to-paraffin' sections (i.e. the original frozen sections were translated to the paraffin domain).}
	\label{fig:examples}
\end{figure}

\paragraph{Bag-of-Features Approach}
The CNN features from the extracted patches of the training data were employed for training a k-means clustering model. The number of cluster centers was varied ($c \in \{16, 32, 64, 128\}$). As distance metric, the Euclidean distance was applied.
Based on the cluster model, training and test feature vectors were clustered according to the closest distance to a cluster center and finally aggregated into one histogram for each whole slide image. The histograms constitute the final whole slide image-level feature vector for classification.

\paragraph{Classification Model}
As classification model, we considered a linear SVM as well as an RBF kernel SVM. The cost factor $c$ was optimized in inner cross validation ($c \in \{0.1, 1, 10, 100, 1000\}$). These rather basic models are used due to the small amount of training data needed. Two different approaches (a linear and a more flexible RBF approach) were here explicitly desired, in order to gain further insight.

\begin{figure}[H]
	\centering
	\includegraphics[width=0.9\linewidth]{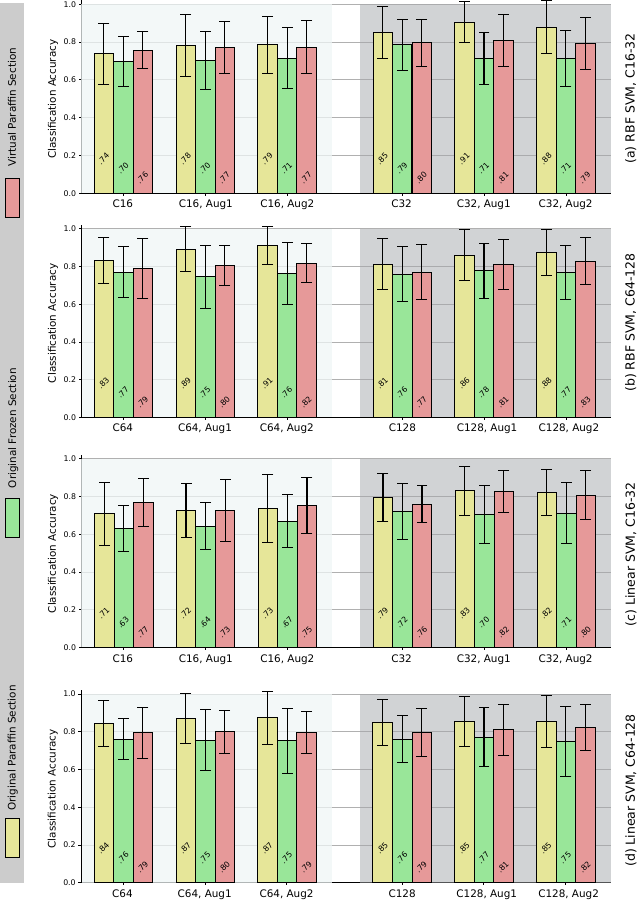}
	\caption{Classification accuracies for different settings consisting of the number of histogram bins (C16, C32, C64, C128) and the data augmentation strategy (default no augmentation and the augmentation strategies Aug1 and Aug2) and the classification model.}
	\label{fig:results}
\end{figure}

\section{Results} \label{sec:results}
Quantitative results of the classification pipeline are shown in Fig.~\ref{fig:results}. For both classification models ((a,b) RBF kernel SVM and (c,d) linear SVM), three augmentation strategies (Aug1, Aug2, default (no augmentation)) and four different clustering strategies (C16, C32, C64, C128), as well as the mean accuracies including the standard deviations, are presented.
While original paraffin sections exhibited the best scores in 32 out of 36 configurations, frozen-to-paraffin translation led to improvements, compared to original frozen sections, in 100 \% of configurations.
The kernel SVM's scores were higher in 24 out of 36 configurations and also the maximum accuracy was higher (0.91 vs 0.87) with this classifier.
Optimum clustering strategy varied, depending on the classifier and the imaging modality. Highest scores for paraffin (and overall) were achieved with C32 and C64. With frozen sections (and also frozen-to-paraffin translations), C128 exhibited the best performance. 
Data augmentation improved the accuracy in 14 out of 16 cases with paraffin sections, in 6 out of 16 cases with frozen section and in 12 out of 16 cases with translated sections.

The output of the image translation process is visualized for four example patches in Fig.~\ref{fig:examples}. While the corresponding images ((b), (c)) show similar underlying tissue, the translated data (c) shows enhanced contrast of e.g. nuclei.

\section{Discussion}

The first goal of this work was to study the effect of the underlying section type on the classification accuracy of thyroid cancer. We performed experiments with many different configurations, in order to be able to make general statements and to be able to determine trends and correlations. 
Comparing overall classification accuracies of original paraffin and frozen section, we noticed that paraffin embedding exhibits the most appropriate section type for automated processing. Superior scores in case of paraffin sections are obtained completely independently of the underlying image analysis models' configuration. Based on these outcomes, we conclude that the lower visual quality of frozen section obviously not only affects manual assessment, but also computer-based decision support techniques. 
Accuracy of computer vision approaches, obviously similarly suffers from the less appropriate modality even though there is not drift between distribution considering training and test data. 
Potentially, if also the CNN feature extraction stage is adapted (and not only pre-trained on a different data set), this effect could be diminished. However, recent work indicates that also the performance of deep learning-based approaches depends on the quality of visual features~\cite{Gadermayr19b}.
As variability of frozen sections is typically higher, the effect is expected to be even stronger in case of multi-centered studies.

When translating the frozen section to paraffin, we noticed a clear improvement of accuracies. This effect was also prevalent, independent of the setting, but was particularly strong for low dimensional histograms (C16, C32) and also if a linear classification model was used. We expected that the image translation approach not only translates the data from one modality to the other, but also performed an effective kind of normalization (see also Fig.~\ref{fig:examples}). This assumption was also confirmed by the fact that with specific settings (C16, linear SVM), the scores obtained with original paraffin sections were even outperformed. 

The data augmentation strategy presented here improved scores in the majority of configurations with paraffin and frozen-to-paraffin sections. In the augmentation approach, there was a trade-off between diversity of data and inaccuracy in the distribution due to a reduced number of patches per histogram (in the bag-of-features approach). We assume that the negative performance of the data augmentation approach in case of original frozen sections is caused by the higher level of variability within the individual whole slide images. For both approaches, dealing with real or fake paraffin sections, one of the augmentation strategies exhibits the best configuration. Finally, there is also a trend that Aug2 with fewer samples per whole slide image and thereby higher variability in the training data, performs best with a higher number of cluster centers (e.g. C128). This is obviously due to the fact that the higher number of features require a higher variability in the training data.

\section{Conclusion}
In this work, we showed that the reduced image quality of frozen sections, which limits accuracy for pathologists, also leads to a clearly reduced classification accuracy of an automated classification system. This effect was observed widely independently of the underlying image analysis model.
We further showed that image translation, from frozen to paraffin sections, based on a state-of-the-art generative adversarial network, was able to clearly increase the performance in case of frozen sections and reduced the gap to real paraffin sections. The proposed data augmentation strategy increased the scores of virtual paraffin sections in the classification setting even further.
Finally, this work provides a strong motivation for performing a study with expert pathologists performing categorization of frozen sections and the corresponding translated sections, to investigate whether a similar effect is achieved in a clinical setting.

\subsection*{Acknowledgement} 
This work was partially funded by the County of Salzburg under grant number FHS-2019-10-KIAMed.




\bibliographystyle{splncs04}

\bibliography{bib}

\end{document}